\documentclass[%
aps,
pra,
amsmath,amssymb,
preprint,%
superscriptaddress,
floatfix
]{revtex4-2}

\usepackage{dcolumn}%
\usepackage{bm}%
\usepackage[utf8]{inputenc}
\usepackage[T1]{fontenc}
\usepackage{mathptmx}
\usepackage{etoolbox}
\usepackage{physics}
\usepackage{graphicx}%
\graphicspath{./ }
\usepackage{amsmath}
\usepackage[dvipsnames]{xcolor}
\usepackage[colorlinks=true,
citecolor=blue,
linkcolor=magenta]{hyperref}

\begin{document}

\title{1-GHz VIS-to-MIR frequency combs enabled by CMOS-compatible nanophotonic waveguides}

\author{Xuan Zhang}
    \affiliation{Shanghai Institute of Optics and Fine Mechanics, Chinese Academy of Sciences, Shanghai 201800, China}
    \affiliation{University of Chinese Academy of Sciences, Beijing 100049, China}
\author{Yuchen Wang}
    \email{wangyuchen@siom.ac.cn}
    \affiliation{Shanghai Institute of Optics and Fine Mechanics, Chinese Academy of Sciences, Shanghai 201800, China}
\author{Junguo Xu}
    \affiliation{Shanghai Institute of Optics and Fine Mechanics, Chinese Academy of Sciences, Shanghai 201800, China}
    \affiliation{University of Chinese Academy of Sciences, Beijing 100049, China}
\author{Qiankun Li}
    \affiliation{Key laboratory of Specialty Fiber Optics and Optical Access Networks, Shanghai University, Shanghai 200443, China}
\author{Xueying Sun}
\author{Yongyuan Chu}
    \affiliation{Key laboratory of Specialty Fiber Optics and Optical Access Networks, Shanghai University, Shanghai 200443, China}
\author{Xiyue Zhang}
    \affiliation{Shanghai Institute of Optics and Fine Mechanics, Chinese Academy of Sciences, Shanghai 201800, China}
    \affiliation{University of Chinese Academy of Sciences, Beijing 100049, China}
\author{Xia Hou}
    \affiliation{Shanghai Institute of Optics and Fine Mechanics, Chinese Academy of Sciences, Shanghai 201800, China}
\author{Chengbo Mou}
    \affiliation{Key laboratory of Specialty Fiber Optics and Optical Access Networks, Shanghai University, Shanghai 200443, China}    
\author{Hairun Guo}
    \email{hairun.guo@shu.edu.cn}
    \affiliation{Key laboratory of Specialty Fiber Optics and Optical Access Networks, Shanghai University, Shanghai 200443, China}
    \affiliation{Hefei National Laboratory, University of Science and Technology of China, Hefei 230088, China}
\author{Sida Xing}
    \email{xingsida@siom.ac.cn}
    \affiliation{Shanghai Institute of Optics and Fine Mechanics, Chinese Academy of Sciences, Shanghai 201800, China}

\date{\today}

\begin{abstract}
A fully stabilized frequency comb is essential for precision metrology and coherent optical synthesis. However, fully-stabilized frequency combs generally require separate stages for supercontinuum generation  (SCG) and self-referencing, largely limiting their compactness. Here, enabled by the low-threshold multi-octave supercontinuum generation and concurrent third-harmonic generation in low-loss silicon nitride waveguides, we present a novel approach to a self-referenced frequency comb source at 1 GHz repetition rate spanning from the full visible (VIS) to the mid-infrared (MIR). Our coherent comb is seeded by an all-polarization-maintaining ultrafast fiber laser at 1556 nm, with a pulse duration of 73 fs at 1 GHz repetition rate. With an injected energy of merely 110 pJ, the pulses propagate through dispersion-engineered Si\textsubscript{3}N\textsubscript{4} waveguides , generating supercontinuum spanning over three octaves from 350-3280 nm i.e. 0.76 PHz of coherent bandwidth. Moreover, the on-chip third harmonic (TH) generation provides a carrier envelope offset beat note via \textit{f-3f} with a signal-to-noise ratio of 43 dB. Fueled by the evolving photonic integration providing possibilities of on-chip filtering and photo-detectors, this approach for single-chip self-referencing of high-repetition-rate frequency combs paves the way for ultrabroadband comb sources with unprecedented compactness and field-readiness.

\end{abstract}
\maketitle
\section{Introduction}
Broadband optical frequency combs underpin a wide range of applications \cite{picque2019,newbury2011,diddams2010}, including high resolution timing and synchronization \cite{8,9}, molecular spectroscopy \cite{18,2,10,14,13}, and environmental sensing \cite{refs-40,refs-41}. Recent demonstrations of gigahertz-rate combs have enabled fast hyperspectral imaging of biological samples \cite{refs-42}, sub-micrometer precision in dual-comb LiDAR \cite{refs-27}, and stable wavelength calibration in astronomical spectrographs \cite{4,6}. These capabilities reflect a broader trend toward faster acquisition rates, finer spectral resolution, and long-term operational stability. Comb sources with gigahertz repetition rates and broadband output have thus become increasingly important for scalable, high-performance frequency referencing across emerging platforms.

Ultrafast fiber lasers \cite{fermann2013} offer a uniquely favorable combination of coherence and stability, making them ideal seed sources for broadband comb generation. Their pulse characteristics are well understood and supported by established models, enabling predictable spectral shaping and offset detection across nonlinear platforms. These lasers have been successfully used to drive SCG in highly nonlinear fibers \cite{refs-36,lesko2020,hutter2025}, lithium niobate \cite{refs-28}, Aluminum nitride \cite{30}, gallium nitride \cite{refs-31}, and silicon nitride waveguides \cite{refs-32}. Their mechanical robustness and long-term coherence support both laboratory and field deployment, while their compatibility with existing photonic components facilitates scalable integration. These attributes make fiber-based lasers a natural choice for initiating multi-octave comb generation in compact, high-repetition-rate platforms.

Despite the maturity of broadband comb generation at low repetition rates, a fully integrated architecture that simultaneously supports gigahertz repetition rate, broadband output, and direct offset detection remains unavailable. At higher repetition rates, reduced pulse peak powers constrain nonlinear interactions, which makes spectral broadening and offset detection significantly more difficult. Existing gigahertz comb platforms, including fiber-based lasers \cite{refs-28,refs-31,lesko2020}, Kerr microresonators \cite{refs-30}, electro-optic combs \cite{20}, and SESAM mode-locked lasers \cite{refs-29,refs-38}, typically offer only single-octave bandwidth and rely on external stabilization. While each platform presents distinct advantages, current implementations remain limited in their ability to simultaneously support broadband output, gigahertz repetition rate, and direct offset detection within an integrated system. Other systems have extended spectral coverage beyond 3 $\mu$m by combining SCG in highly nonlinear fibers with intra-pulse difference frequency generation using nonlinear crystals or cascaded media \cite{21,refs-37}. While effective, these approaches introduce significant complexity and remain incompatible with full integration. Maintaining nonlinear efficiency at gigahertz repetition rates while achieving broadband output and offset detection within an integrated platform remains an open challenge. This constraint continues to hinder the realization of broadband gigahertz combs in scalable, self-referenced systems.

In this work, we break this limitation by demonstrating a fully self-referenced frequency comb operating at 1 GHz repetition rate. Using dispersion-engineered silicon nitride waveguides, we access a previously inaccessible regime of multi-octave spectral broadening and third-harmonic generation with only 110 pJ pulse energy. The system spans over three octaves (350 nm–3280 nm) and provides an on-chip Carrier envelope offset (CEO) beat note with 43 dB signal-to-noise ratio. We demonstrate a GHz-rate comb source that achieves multi-octave bandwidth and on-chip self-referencing without external amplification or free-space optics, offering a scalable and field-deployable solution for integrated photonic platforms.

\begin{figure*}[!ht]
\centering\includegraphics[width=\linewidth]{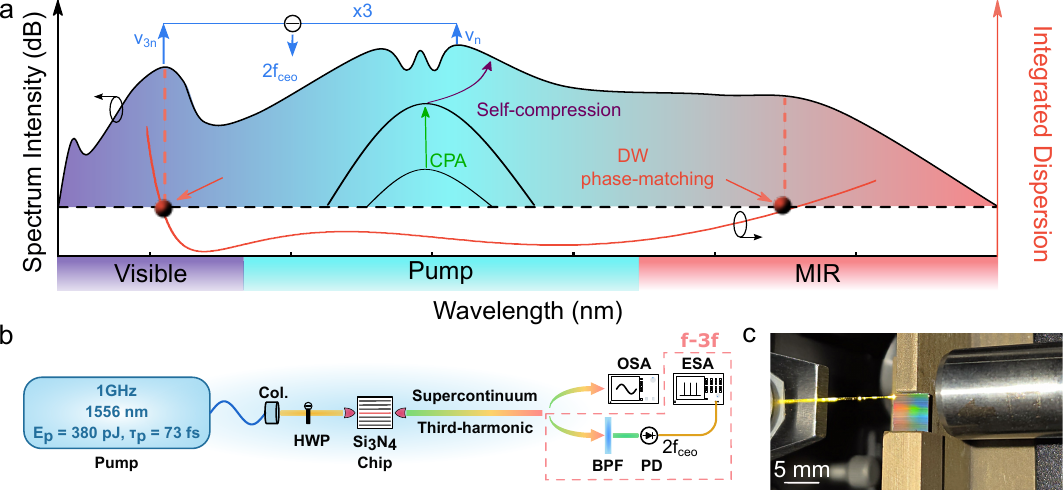}
\caption{\textbf{Principle and experimental setup of the multi-octave MIR comb generation.}
a Illustration of physical processes for on-chip one-step self-referencing. CPA: Chirped-pulse amplification; DW: Dispersive wave.
b Schematic setup of SC generation with 1-GHz all-fiber optical frequency comb. Col.: Collimator; HWP: Half-wave plate; OSA: Optical spectrum analyzer; ESA: Electrical spectrum analyzer; PD: Photodetector; BPF: Band-pass filter.
c Photograph of the waveguide generating visible light. 
}
\label{fig1}
\end{figure*}

\section{All-polarization-maintaining 1-GHz fs fiber laser system}
\begin{figure*}[!ht]
\includegraphics[width=\linewidth]{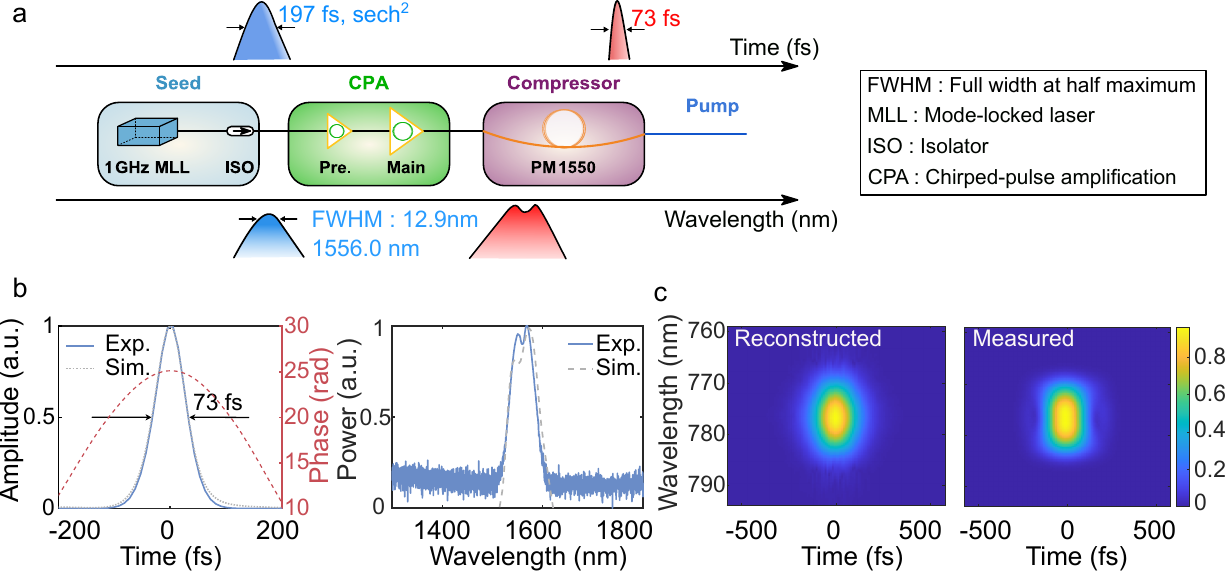}
\caption{\label{fig-1GHzpump}\textbf{Characterization of the 1 GHz femtosecond pump source.}
a Diagram of the 1 GHz femtosecond pump laser setup.
b Left: Experimentally retrieved (blue) and simulated (gray) SHG-FROG traces of the 73 fs pulse; Right: Experimentally recorded spectrum (blue) and simulated (gray) spectrum after self-compression segment.
c Left: Reconstructed second harmonic generation frequency resolved optical gating (SHG-FROG) trace of the compressed pulses; Right: Experimental SHG-FROG trace. }
\end{figure*}

The design and implementation of the 1 GHz ultrafast pump source are detailed below, as it serves as the basis for broadband SCG. A schematic of the complete setup is shown in FIG. \ref{fig-1GHzpump}a. The mode-locked laser is a commercially available model (MENHIR-1550) that generates $\sim$197 fs sech\textsuperscript{2} pulses with 13 nm of bandwidth. To prevent feedback-induced instabilities, an optical isolator is incorporated before coupling into the amplification stages.

The all-fiber amplifier employs a chirped-pulse amplification (CPA) scheme, in which dispersion is solely managed by the gain fibers without any external dispersion-compensating components. Pulses from the mode-locked laser are first amplified in a pre-amplifier stage using an erbium-doped fiber amplifier (EDFA), seeded with 48 mW of power. After initial amplification, the pulses are temporally stretched in dispersion-compensating fiber and subsequently amplified by a second, identical EDFA. The two-stage amplifier is pumped using butterfly-packaged 980-nm laser diodes in each stage to provide controlled gain. The pre-amplifier incorporates $\sim$90 cm of Er-80/4-125-HD-PM fiber, while the main amplifier uses $\sim$120 cm of the same fiber type. Both stages utilize normal-dispersion gain fibers to suppress nonlinear distortions and manage pulse chirping, ensuring minimal phase error and stable pulse shaping. The final average output power is approximately 380 mW, corresponding to 380 pJ pulse energies.

To obtain sufficient peak power for nonlinear spectral broadening, pulse compression is built by fusion-splicing a segment of PM1550 fiber (D = 18 ps/nm/km) to the second EDFA. The interplay between group velocity dispersion and Kerr nonlinearity in this segment enables pulse self-compression. A cutback experiment, starting from an initial fiber length of 100 cm, was conducted to optimize compression performance. Detailed spectral results from this chirp optimization are presented in Supplementary Fig. 1, where the optimal fused fiber length was identified to be approximately 42 cm. Under this condition, the compressed pulse duration was measured at 73 fs using second-harmonic generation frequency-resolved optical gating (SHG-FROG), with the retrieved and reconstructed traces shown in FIG. \ref{fig-1GHzpump}b. The retrieved spectrogram is in agreement with the reconstructed one, confirming a pulse duration of about 73 fs and thus a peak power of $\sim$5 kW.

This polarization-maintaining all-fiber MOPA, based on CPA amplification and self-compression, minimizes system complexity and eliminates the need for free-space dispersion compensation or compression components, thereby favoring compact and robust implementation for on-chip nonlinear photonic applications. All fiber components used in the laser and amplification system are polarization-maintaining, ensuring stable operation and alignment-free integration.

\section{VIS-to-MIR SCG with SiN photonic chip}
In Si\textsubscript{3}N\textsubscript{4} waveguides, MIR spectral components are primarily generated through soliton dynamics and dispersive wave (DW) emission, governed by engineered dispersion profiles. The high Kerr nonlinearity and tight mode confinement support efficient spectral broadening. DW emission typically arises near the zero-dispersion wavelength (ZDW), often located in the near-infrared, while soliton self-frequency shift (SSFS) can extend the spectrum toward the MIR. Compared to silica fibers, SSFS in Si\textsubscript{3}N\textsubscript{4} is constrained by its relatively low Raman gain and tailored dispersion characteristics. Moreover, higher-order dispersion and waveguide losses in the MIR regime further limit red-shifted extension. Nevertheless, with precise dispersion engineering and controlled input pulse chirp, coherent MIR generation is attainable with preservation of the comb structure. In our experiments, spectral broadening was primarily characterized under transverse electric (TE) mode excitation, controlled via a half-wave plate. Comparative measurements under transverse magnetic (TM) mode excitation showed significantly reduced MIR extension, as detailed in the Supplementary Information.

\subsection{Waveguide Dispersion Engineering for Controlled Spectral Broadening}
\begin{figure*}[!ht]
    \centering  
    \includegraphics[width=\linewidth]{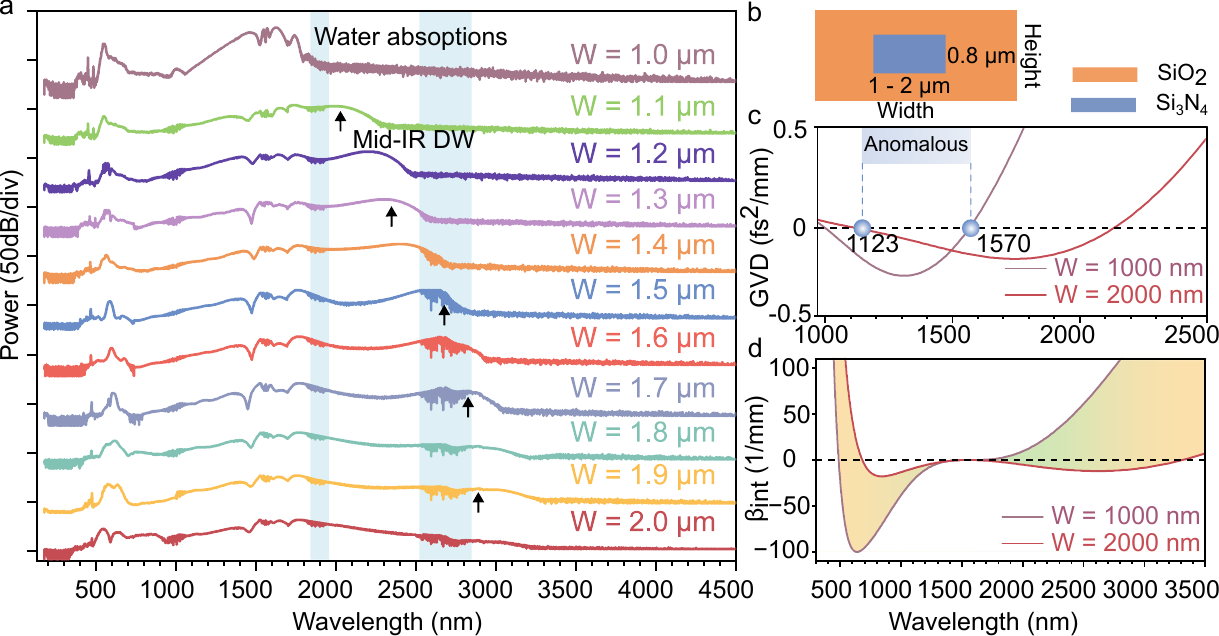}
    \caption{
   \textbf{Dispersion-engineered waveguides for supercontinuum generation.}
    a Experimental SC spectra for different waveguide widths. 
    b The schematic of the waveguide structure.
    c Simulated Group velocity dispersion of 800 nm-thick Si$_3$N$_4$ waveguides.
    d Simulated integrated dispersion of 800 nm-thick Si$_3$N$_4$ waveguides with core widths ranging from 1000 nm (red) to 2000 nm (orange).}
    \label{fig-disperisonSC}
\end{figure*}

The Si\textsubscript{3}N\textsubscript{4} waveguides used in this work are fabricated using the photonic Damascene process \cite{34}, which allows for a deposited Si\textsubscript{3}N\textsubscript{4}  film as thick as $\sim$1 {\textmu}m. The waveguides feature a SiO\textsubscript{2} core with Si\textsubscript{3}N\textsubscript{4} cladding, supporting high mode confinement and enabling strong nonlinear interactions. The schematic of this structure is shown in FIG. \ref{fig-disperisonSC}b. Mode profiles and dispersion characteristics were numerically simulated using the finite element method (COMSOL), with FIG. \ref{fig-disperisonSC}c and FIG. \ref{fig-disperisonSC}d showing the GVD and integrated dispersion for waveguide widths from 1000 to 2000 nm. Lithographic tuning of the waveguide width (w = 1000-2000 nm, h = 800 nm) provides precise dispersion engineering, as confirmed by numerical simulations of the group-velocity dispersion (GVD), shown in FIG. \ref{fig-disperisonSC}c. At a pump wavelength of 1556 nm, the waveguides operate in the anomalous dispersion regime, enabling soliton formation and nonlinear spectral broadening.  

To analyze phase matching for DW emission, we computed the integrated dispersion as:
\begin{equation} \label{eq:1}
\beta_{\mathrm{int}}(\omega)=\sum_{k\geq2}\frac{(\omega-\omega_s)^k}{k!}\frac{\partial^k\beta}{\partial\omega^k}
\end{equation}
\noindent where $\omega_s$ is the soliton central frequency. $\beta_{int}$ provides insight into how HOD influences the spectral positioning and efficiency of DW generation. By tailoring the waveguide geometry to control $\beta_{int}$, we achieve efficient phase matching in both mid-infrared and visible regions. This interplay of soliton dynamics and dispersion design finally underpins the generation of supercontinua extending beyond 3 {\textmu}m. 

With the waveguide width varying from 1000 to 2000 nm, dispersion profiles and phase-matching conditions are tailored to control the soliton-dispersive wave dynamics and to boost spectral extension into the mid-infrared. FIG. \ref{fig-disperisonSC} shows the SC spectra measured across the waveguide width range, revealing systematic shifts in dispersive wave positions. With increasing waveguide width, both short- and long-wavelength dispersive waves redshift, shifting phase-matching toward longer wavelengths and driving SC extension into the mid-infrared. These trends are consistent with the simulated integrated dispersion profiles in FIG. \ref{fig-disperisonSC}d, which show that the positions of phase-matched dispersive waves closely follow the evolution of the ZDWs with increasing width. Among all tested structures, the 1900 nm-wide waveguide yielded the broadest spectrum, demonstrating an optimal balance between dispersion phase-matching and nonlinear interaction strength \cite{31}. 

Notably, despite the presence of multiple ZDWs and an extended anomalous dispersion regime in the 2000 nm-wide waveguide, the observed spectrum fails to reach deeper into the mid-infrared. This deviation from the expected behavior may be attributed to insufficient on-chip pulse energy, which falls below the nonlinear threshold required for efficient soliton fission and inhibits further redshifted dispersive wave generation \cite{28}, or to increased losses in the SiO\textsubscript{2} cladding for the mid-infrared radiation. The estimated total insertion efficiency from the pump beam into the waveguide and then measured after the output aspheric lens is approximately 30$\%$, which limits the injected pulse energy and contributes to this spectral suppression. It was also noted that the measured SC features water absorption lines in the range of 1.8-1.9 {\textmu}m and 2.5-2.8 {\textmu}m, which occur in the path towards and within the FTIR device (FIG. \ref{fig-disperisonSC}).

\subsection{Pre-Chirping for Spectral Tailoring}
With the broadest SC achieved in the 1900-nm-wide waveguide, whose shortwave DW is located at 630 nm, the chirp of the input pulse is subsequently tuned to control soliton dynamics and DW generation, further enhancing the MIR coverage. In Si\textsubscript{3}N\textsubscript{4} waveguides, where Raman effects are negligible and the SSFS is suppressed, the role of pre-chirping shifts from controlling red-shifted soliton dynamics to modulating the temporal compression and the fission of the injected soliton. A negative chirp enhances peak power within the waveguide, facilitating earlier soliton fission and stronger DW emission. Consequently, pre-chirping primarily governs the efficiency and spectral reach of short-wavelength components through its influence on soliton-driven DW generation.

\begin{figure}[!ht]
    \centering  
    \includegraphics[width=\linewidth]{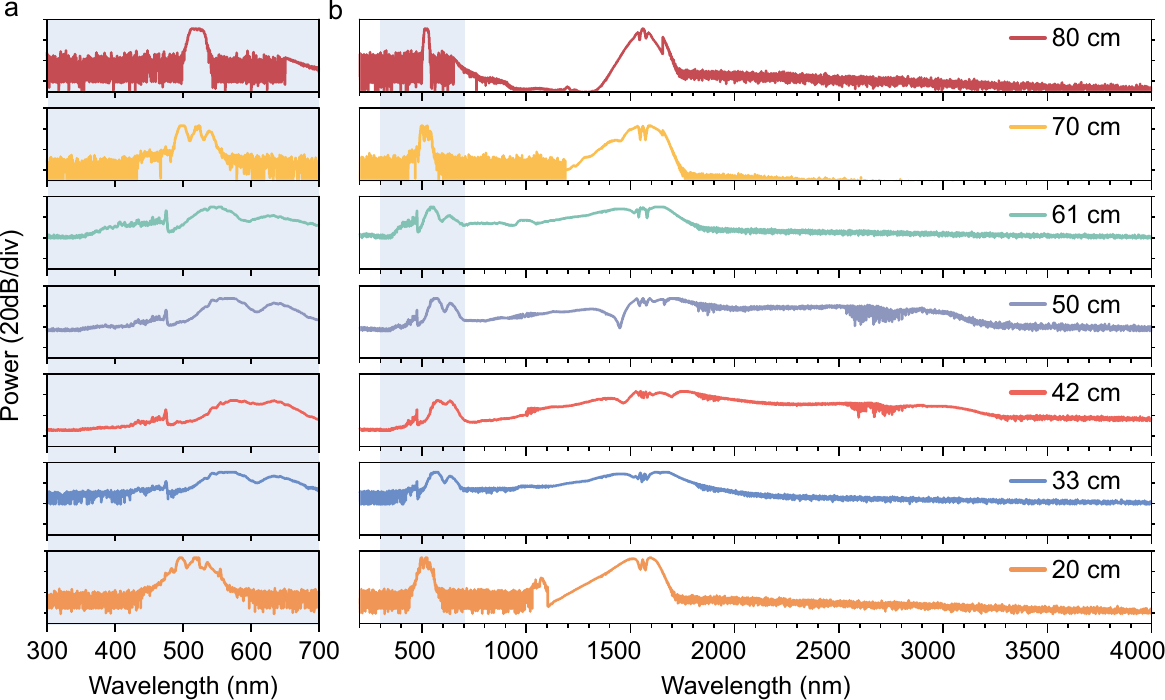}
    \caption{
    \textbf{Supercontinuum generation under different input conditions.}
    a zoom-in visible spectra (500–700 nm) collected at the output of a 1900 nm-wide waveguide under varying pre-chirp conditions.
    b Corresponding full supercontinuum spectra spanning the entire measurement range.
}
    \label{fig-prechirp}
\end{figure}

Building on the initial cut-back experiment, a larger step size was selected to systematically study the effect of pre-chirp control on the SCG and in particular the MIR DW. Fiber lengths ranging from 80 cm to 33 cm were evaluated to systematically vary the amount of the pre-chirp. At the optimal length of 42 cm, a maximal spectral coverage was observed. Deviations from this condition, either toward positive or excessive negative chirp led to reduced bandwidths, which may due to the reduced peak power at the waveguide input, thereby suppressing soliton-driven dynamics and limiting both short- and long-wavelength extension. As shown in FIG.\ref{fig-prechirp}a, the resulting SC spectrum evolves significantly with fiber length. At longer lengths (e.g., 80 cm), the input pulses exhibit strong positive chirp, resulting in reduced peak power and minimal nonlinear broadening, with spectral components largely confined to the fundamental and weak TH generation. As the fiber length is reduced, temporal compression at the waveguide input improves, leading to stronger nonlinear interactions and broader SC generation. The widest spectrum is obtained near 42 cm, indicating optimal dispersion compensation and soliton fission conditions. Further reduction in fiber length introduces negative chirp, which suppresses spectral broadening, and at 33 cm, the SC extends only to $\sim$2800 nm. We observed similar chirp-dependent dynamics and spectral behavior in the 1600 nm-wide waveguide (see Supplementary Information).

\subsection{Numerical modelling}
\begin{figure}[!ht]
    \centering  
    \includegraphics[width=\linewidth]{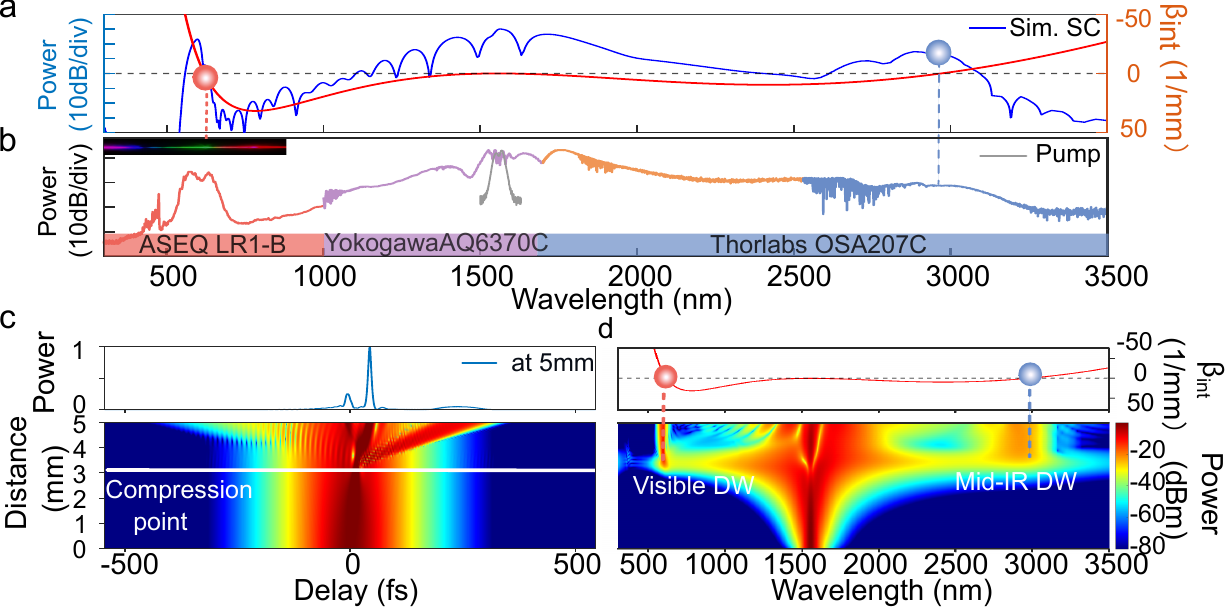}
    \caption{\textbf{Numerical results.}   
    a GNLSE-based simulation of SCG in a SiN waveguide (width: 1900 nm, height: 800 nm), showing SC spectrum (blue) and integrated dispersion profile (red), consistent with experimental data in b.
    b Experimentally measured SC spectrum from the same waveguide, obtained using three complementary instruments for full spectral coverage. The input spectrum is shown by the gray line. 
    c Temporal dynamics of the pulse: simulated output envelope (top) and pump pulse evolution during propagation (bottom).
    d Simulated spectral evolution over the 5 mm waveguide length: integrated dispersion profile (top) and frequency-domain evolution of the pulse spectrum (bottom).
}
    \label{fig-SCsim}
\end{figure}

To demonstrate the full spectral potential of the dispersion-engineered Si\textsubscript{3}N\textsubscript{4} platform under optimized pump conditions, we recorded the broadest achievable SC using the 1900 nm wide waveguide and a pre-chirped pulse compressed by 42 cm of PM1550 fiber. The broadest measured spectrum, shown in FIG. \ref{fig-SCsim}b, spans 350 to 3280 nm or 91.4 - 856.6 THz , encompassing more than 3 octaves or 0.76 PHz of bandwidth.

To analyze the process of the nonlinear spectral broadening in dispersion-engineered Si\textsubscript{3}N\textsubscript{4} waveguides, numerical simulations are performed based on the generalized nonlinear Schrödinger equation (GNLSE) \cite{31}. The Raman effect is negligible due to its weak contribution in Si\textsubscript{3}N\textsubscript{4} waveguides, and considering the short waveguide length of 5 mm, propagation loss was set as $\alpha$ = 0. The model incorporates higher-order dispersion, Kerr nonlinearity, and a self-steepening term. 

Spectral characterization was performed using three complementary spectrometers: ASEQ-LR1-B (200-1200 nm, free-space coupled), Yokogawa AQ6370C (600-1700 nm, fiber-coupled), and Thorlabs OSA207C (1-12 {\textmu}m), the latter used with 1650-nm (orange line in Fig. \ref{fig-SCsim}b and 2400-nm (blue line in Fig. \ref{fig-SCsim}b) long-pass filters (Edmund Optics) to avoid saturation of the Fourier-transform infrared (FTIR) spectrometer by the near-infrared signals. Spectral data from all three instruments are stitched together to obtain the complete supercontinuum profile. Moreover, the spectral coherence of the SC spectrum is numerically evaluated by introducing quantum noise (one photon per mode) into the pump pulses.  

The temporal evolution corresponding to the spectral dynamics shown in FIG. \ref{fig-SCsim}d is presented in Fig. \ref{fig-SCsim}c. After the compression point at approximately 3 mm, the pulse undergoes significant temporal compression followed by splitting, accompanied by DW emission. The emitted DWs rapidly walk off from the residual pump envelope due to their distinct group velocities. Notably, the visible DW appears on the trailing edge of the temporal window, resulting from its increased group delay relative to the pump residue.

\section{On-chip f-3f self-referencing}
To stabilize optical frequency combs and enable precision applications in spectroscopy, metrology, and communications, accurate measurement of f$_\mathrm{ceo}$ is indispensable. Building upon the previous investigation of pre-chirp modulation and waveguide dispersion engineering, we consistently observed strong short-wavelength components around 555 nm in the SC output. To isolate a stable short-wavelength output, the pump power was deliberately reduced to suppress SCG, ensuring that only third-harmonic generation occurred within the waveguide. Under these conditions, a stable third-harmonic component near 555 nm was observed.

\begin{figure*}[!ht]
\centering\includegraphics[width=.8\linewidth]{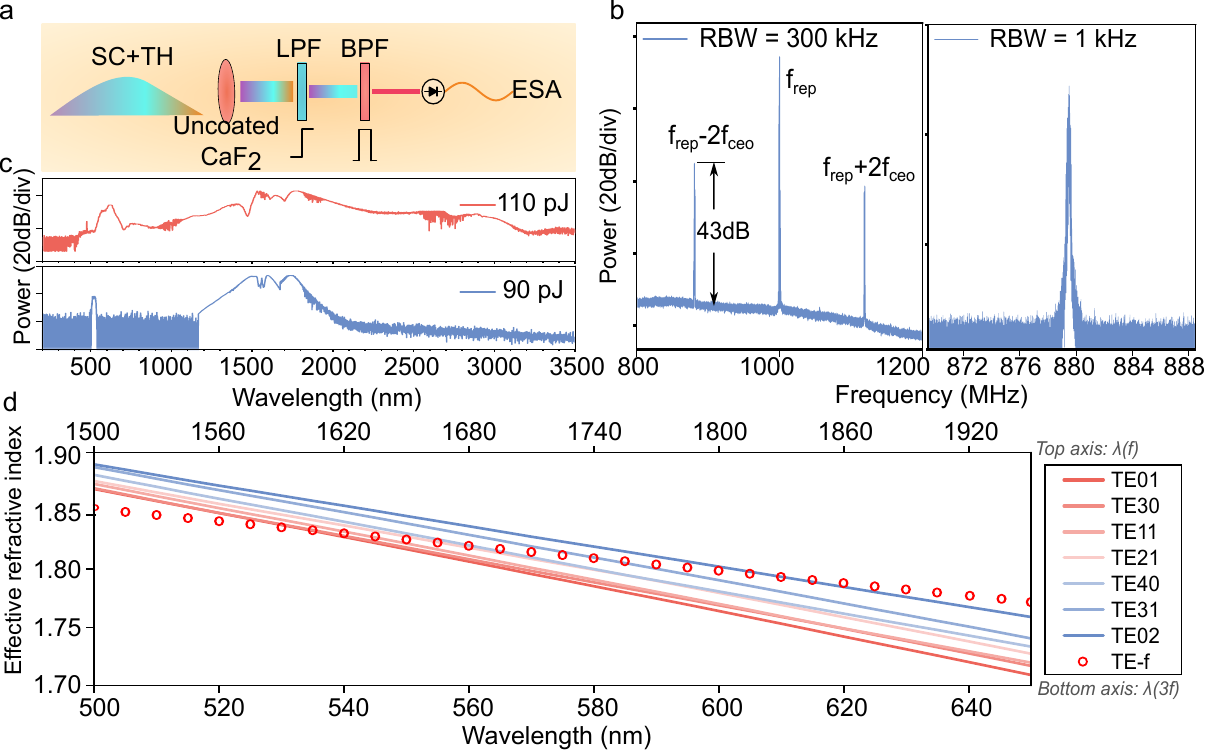}
\caption{\textbf{f-3f self-referencing.}
a Schematic of the f-3f self-referencing setup. 
b Measured beat note signals. 
  Left: $f_{\text{rep}} - 2f_{\text{ceo}}$, $f_{\text{rep}}$ and $f_{\text{rep}} + 2f_{\text{ceo}}$ at 300 kHz RBW.
  Right: $f_{\text{rep}} - 2f_{\text{ceo}}$ at 1 kHz RBW; 
c Output spectra from a 1800 nm wide waveguide under variable launch powers.
d Phase-matching simulation. Fundamental (circles) and third-harmonic (line) mode indices in Si\textsubscript{3}N\textsubscript{4} waveguide, Wavelengths for f and 3f components are shown on top and bottom axes, respectively.
}
\label{fig-f3f}
\end{figure*}

The experimental setup is illustrated in FIG. \ref{fig-f3f}a. For coupling the visible light out of the waveguide, we employed an uncoated calcium fluoride lens. A low-pass optical filter was used to suppress residual pump, followed by a band-pass filter that selectively passed the beat signal to an avalanche photodiode detector (APD210). The resulting radio frequency signal is measured by an electrical spectrum analyzer (ESA) (FIG. \ref{fig-f3f}b). By adjusting the dispersion properties of the waveguide, we achieved a CEO signal with an SNR of 43 dB, using a waveguide width of 1800 nm and a resolution bandwidth of 300 kHz. This provides a robust platform for GHz comb stabilization via visible-band heterodyne detection.

Phase-matching simulations based on effective refractive index (FIG. \ref{fig-f3f}d) confirm optimal modal overlap between fundamental and third-harmonic waves around 555 nm, with the calculated phase-matching window aligning with the experimentally observed THG spectral region. The pump pulses extended into the blue via DW generation in the Si\textsubscript{3}N\textsubscript{4} waveguide, overlapping spectrally with third-harmonic radiation produced by Kerr nonlinearities. This overlap yielded a heterodyne beating signal suitable for $f_{\text{ceo}}$ detection. 

\section{Conclusion}
In conclusion, we demonstrate a fiber-chip hybrid integrated frequency comb source operating at a 1 GHz repetition rate coherently covering a 0.76-PHz bandwidth from visible to MIR. Through GNLSE-based simulations, we quantitatively optimized the fiber length to ensure efficient pulse self-compression under GHz-rate pumping. The pulses are injected into dispersion-engineered Si\textsubscript{3}N\textsubscript{4} waveguides featuring numerically tailored dispersion profiles to facilitate broadband soliton fission and phase-matched dispersive wave generation. Our configuration yields coherent SC spanning more than 3 octaves from 350 nm to 3280 nm i.e. a 0.76-PHz bandwidth, from the ultraviolet into the MIR. Rigorous numerical modeling confirms the solitonic origin of the short-wavelength dispersive wave near 555 nm, which enables direct on-chip f-3f self-referencing. The extracted $f_{\text{ceo}}$ signal achieves an SNR of 43 dB with a resolution bandwidth of 300 kHz, without additional nonlinear components, thus reducing system complexity and favoring compact integration. We demonstrate a visible-to-MIR fiber-chip hybrid integrated frequency comb source beyond GHz rates with on-chip f-3f self-referencing. Moreover, by employing a lens fiber for input coupling and a ZBLAN fiber for output coupling, we further implement an all-fiber–chip hybrid MIR comb (see Supplementary Information for details). Although the present performance is limited, this approach offers a promising pathway toward fully fiberized and chip-integrated MIR comb systems,  paving the way to compact hybrid-integrated high-repetition-rate frequency combs suitable for broadband multi-species molecular fingerprinting \cite{13,14}, breath diagnostics \cite{refs-43}, and astronomical spectrograph calibration \cite{4,5,6}. 

\section*{Supplementary information}
See the supplementary material for supporting content.

\bibliography{1Gmain}

\end{document}

\title{Supplementary Information for: 1-GHz VIS-to-MIR frequency combs enabled by CMOS-compatible nanophotonic waveguides}

\author{Xuan Zhang}
    \affiliation{Shanghai Institute of Optics and Fine Mechanics, Chinese Academy of Sciences, Shanghai 201800, China}
    \affiliation{University of Chinese Academy of Sciences, Beijing 100049, China}
\author{Yuchen Wang}
    \email{wangyuchen@siom.ac.cn}
    \affiliation{Shanghai Institute of Optics and Fine Mechanics, Chinese Academy of Sciences, Shanghai 201800, China}
\author{Junguo Xu}
    \affiliation{Shanghai Institute of Optics and Fine Mechanics, Chinese Academy of Sciences, Shanghai 201800, China}
    \affiliation{University of Chinese Academy of Sciences, Beijing 100049, China}
\author{Qiankun Li}
    \affiliation{Key laboratory of Specialty Fiber Optics and Optical Access Networks, Shanghai University, Shanghai 200443, China}
\author{Xueying Sun}
\author{Yongyuan Chu}
    \affiliation{Key laboratory of Specialty Fiber Optics and Optical Access Networks, Shanghai University, Shanghai 200443, China}
\author{Xiyue Zhang}
    \affiliation{Shanghai Institute of Optics and Fine Mechanics, Chinese Academy of Sciences, Shanghai 201800, China}
    \affiliation{University of Chinese Academy of Sciences, Beijing 100049, China}
\author{Xia Hou}
    \affiliation{Shanghai Institute of Optics and Fine Mechanics, Chinese Academy of Sciences, Shanghai 201800, China}
\author{Chengbo Mou}
    \affiliation{Key laboratory of Specialty Fiber Optics and Optical Access Networks, Shanghai University, Shanghai 200443, China}    
\author{Hairun Guo}
    \email{hairun.guo@shu.edu.cn}
    \affiliation{Key laboratory of Specialty Fiber Optics and Optical Access Networks, Shanghai University, Shanghai 200443, China}
    \affiliation{Hefei National Laboratory, University of Science and Technology of China, Hefei 230088, China}
\author{Sida Xing}
    \email{xingsida@siom.ac.cn}
    \affiliation{Shanghai Institute of Optics and Fine Mechanics, Chinese Academy of Sciences, Shanghai 201800, China}

\date{\today}%

\maketitle
\section{Supplementary Spectra of Pre-Chirped Input Pulse}
\begin{figure}[h]
    \centering  
    \includegraphics[width=.6\linewidth]{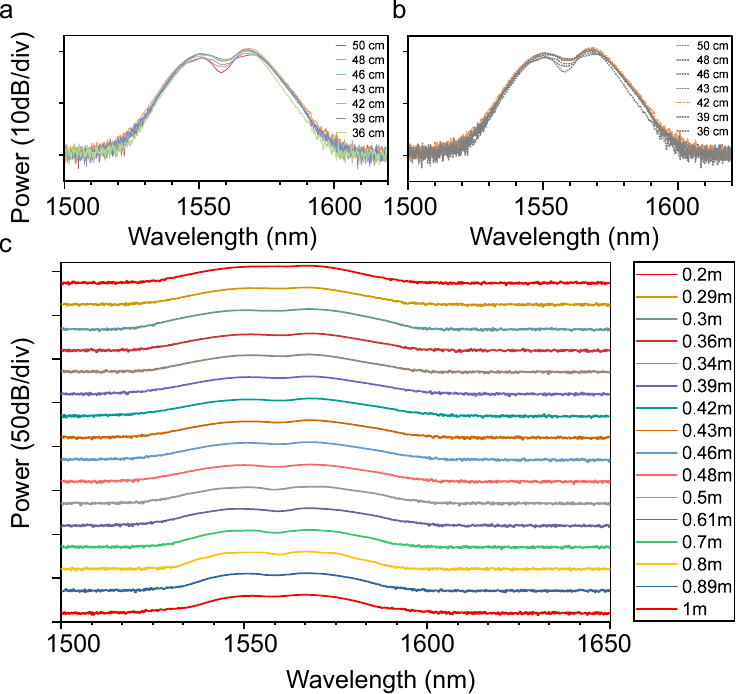}
    \caption{
   a Spectral results from the cutback experiment near the optimal chirp length (36-50 cm).  
   b Overlay of all spectra in gray, with the 42 cm case highlighted in orange to indicate the optimal chirp condition.  
   c Full set of spectra from the cutback experiment, showing the evolution of spectral broadening as the fused fiber length is reduced from 1 m to 0.2m.
}
    \label{fig:prechirp}
\end{figure}

To optimize the input pulse chirp for efficient supercontinuum generation, we performed a cutback experiment by gradually shortening the length of the fused fiber used for pre-chirping the pump pulse. Starting from an initial fiber length of 1 m, the fiber was cut back in steps down to 0.2 m, and the corresponding spectra were recorded. The full set of spectral results is shown in FIG. \ref{fig:prechirp}c. By comparing the experimental data with numerical simulations, the optimal chirp condition was identified at a fiber length of approximately 42 cm, which yielded the broadest and most coherent spectrum. FIG. \ref{fig:prechirp}.a and FIG. \ref{fig:prechirp}.b highlight the spectral evolution near the optimal chirp length (36–50 cm). In FIG. \ref{fig:prechirp}.b, all spectra are plotted in gray except for the 42 cm case, which is shown in orange to emphasize its superior spectral broadening performance.

\section{TM Polarization Measurements in SiN Waveguides}
\begin{figure}[h]
    \centering  
    \includegraphics[width=.7\linewidth]{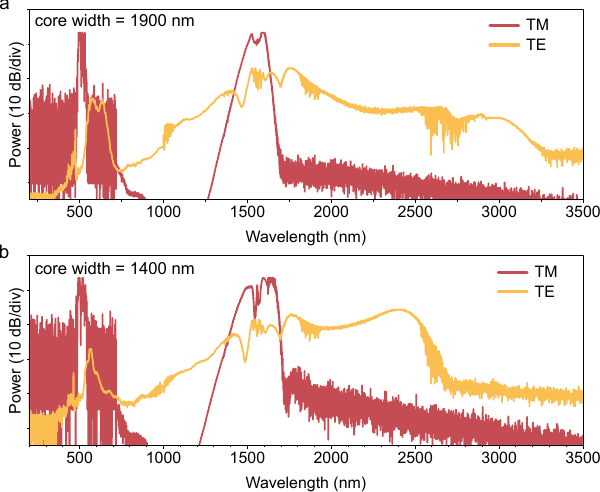}
    \caption{
   Output spectra under TE and TM excitation in Si\textsubscript{3}N\textsubscript{4} waveguides of 1900 nm (a) and 1400 nm (b) widths.
}
    \label{fig:TM}
\end{figure}

Spectral measurements were performed under Transverse Electric (TE) and Transverse Magnetic (TM) excitation in Si\textsubscript{3}N\textsubscript{4} waveguides with widths of 1900 nm and 1400 nm (see FIG. \ref{fig:TM}.). The pump pulses (central wavelength: 1556 nm; duration: 73 fs; pulse energy: 380 pJ) were polarization-controlled using a half-wave plate. TE excitation yielded broadband supercontinuum generation in both waveguides. In contrast, TM excitation led to the appearance of third-harmonic components without noticeable spectral broadening. The absence of continuum features under TM pumping is attributed to reduced mode confinement and less favorable dispersion conditions, which collectively suppress the nonlinear interactions required for broadband generation. Consistent results across both waveguide geometries confirm that this limitation is not structure-specific, but rather intrinsic to TM-mode excitation under the examined conditions.

\section{Supplementary pre-chirp control in 1600 nm-Wide Waveguide}
\begin{figure}[h]
    \centering  
    \includegraphics[width=.7\linewidth]{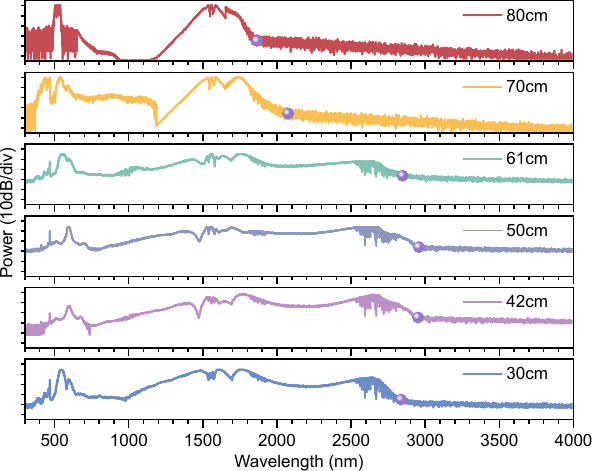}
    \caption{
   Full SC spectra measured at the output of a 1600 nm-wide waveguide under various pre-chirp conditions.
}
    \label{fig:1600nm-prechirp}
\end{figure}

We further investigated the influence of pre-chirp modulation on supercontinuum generation using a Si\textsubscript{3}N\textsubscript{4} waveguide with a width of 1600 nm (see FIG. \ref{fig:1600nm-prechirp}). Compared to the 1900 nm-wide waveguide described in the main text, the spectral bandwidth is reduced due to differences in waveguide geometry. Under identical pre-chirp control conditions, the short-wavelength dispersive wave and third-harmonic generation components exhibit consistent evolution trends. In particular, both features reach their peak response near the transform-limited input condition, confirming that the mechanisms governing short-wavelength generation remain robust across moderate variations in waveguide design.

\section{Towards All-fiber integrated}
\begin{figure}[!ht]
    \centering  
    \includegraphics[width=\linewidth]{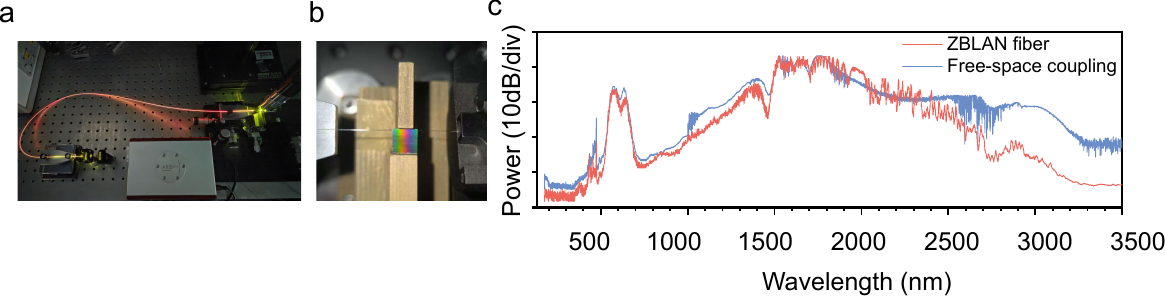}
    \caption{
   a Image of the ZALAB laser output setup. 
   b Fully fiber-integrated system design.
   c Comparative output spectra acquired using ZBLAN fiber coupling (blue line) versus free-space coupling configuration (orange line).
}
    \label{fig:ZBLAN}
\end{figure}

We performed comparative measurements using both ZBLAN fiber and hybrid coupling schemes (see FIG.  \ref{fig:ZBLAN}. In the hybrid setup (blue line), visible components were collected using uncoated calcium fluoride lenses, near-infrared signals were coupled via lens-fiber interfaces, and mid-infrared light was collected using D-coated lenses optimized for longer wavelengths. In contrast, the ZBLAN fiber (red line) provided a single-channel collection approach. 
 
\section{Output Spectra at Reduced Launch Powers}
\begin{figure}[!ht]
    \centering  
    \includegraphics[width=.7\linewidth]{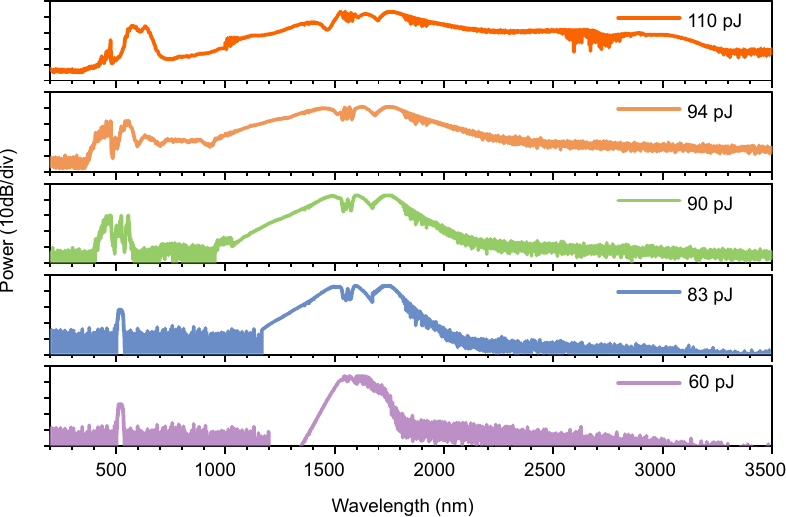}
    \caption{Spectral evolution with decreasing launch power. }
    \label{fig:power}
\end{figure}

To investigate the influence of launch power on spectral broadening, we measured the output spectra as the input pulse energy was gradually reduced from 110 pJ to 60 pJ. As shown in FIG. \ref{fig:power}, the supercontinuum bandwidth narrows progressively with decreasing launch power, indicating reduced nonlinear interaction strength. At approximately 90 pJ, the input energy becomes insufficient to support soliton formation, and no further spectral broadening is observed. Below this threshold, the output spectrum is dominated solely by third-harmonic(TH) generation, with no evidence of dispersive wave emission or continuum extension. This transition highlights the critical role of soliton dynamics in driving broadband generation and confirms the power-dependent onset of nonlinear processes in the Si\textsubscript{3}N\textsubscript{4} waveguide.

\section{Supplementary f-3f Measurements}
\begin{figure}[h]
    \centering  
    \includegraphics[width=.7\linewidth]{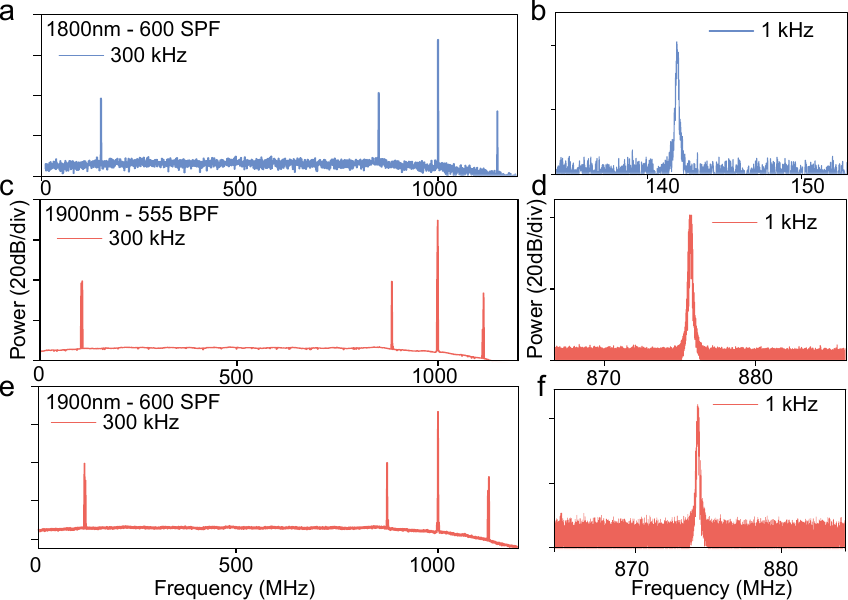}
    \caption{
    f\textsubscript{ceo} beat notes measured via heterodyne detection of short-wavelength components. 
a,b 1800 nm-wide waveguide with 600 nm short-pass filter; 
c,d 1900 nm-wide waveguide with 555 nm band-pass filter;
e,f 1900 nm-wide waveguide with 600 nm short-pass filter. 
a,c,e show results at 300 kHz RBW; 
b,d,f at 1 kHz RBW. 
Blue traces: 1800 nm waveguide; red traces: 1900 nm waveguide. All configurations yield 40 dB SNR.
}
    \label{fig:fceo}
\end{figure}

To further validate the coherence of our GHz comb system, we performed rigorous CEO frequency measurements using heterodyne detection of TH and DW components in 1900 nm-wide Si\textsubscript{3}N\textsubscript{4} waveguides. As shown in FIG. \ref{fig:fceo}, clear f\textsubscript{ceo} beat notes were obtained under both 300 kHz and 1 kHz resolution bandwidths (RBW), with SNR exceeding 40 dB. Three representative configurations are presented: a,b using a 600 nm short-pass filter in an 1800 nm-wide waveguide (blue traces); c,d using a 555 nm band-pass filter in a 1900 nm-wide waveguide (red traces); and e,f using a 600 nm short-pass filter in the same 1900 nm waveguide (red traces). 
 
\section{Coherence Analysis of the Generated Supercontinuum}
The spectral coherence of the SC spectrum is numerically evaluated by introducing quantum noise (one photon per mode) into the pump pulses. The module of the complex degree of first-order coherence is computed using
$|\Gamma^{(1)}(\lambda)|=|\frac{\langle E_1^*(\lambda)E_2(\lambda)\rangle}{\langle|E_1(\lambda)|^2\rangle^{1/2}\langle|E_2(\lambda)|^2\rangle^{1/2}}|$, where ( E$_1$($\lambda$) )and ( E$_2$($\lambda$)) are the complex spectral envelopes of independently generated SC pairs, and the brackets denote ensemble averaging \cite{dudley2002}. As shown in supplement information Fig\ref{fig:coherence}, revealing high coherence. 
 
\begin{figure}[!h]
    \centering  
    \includegraphics[width=\linewidth]{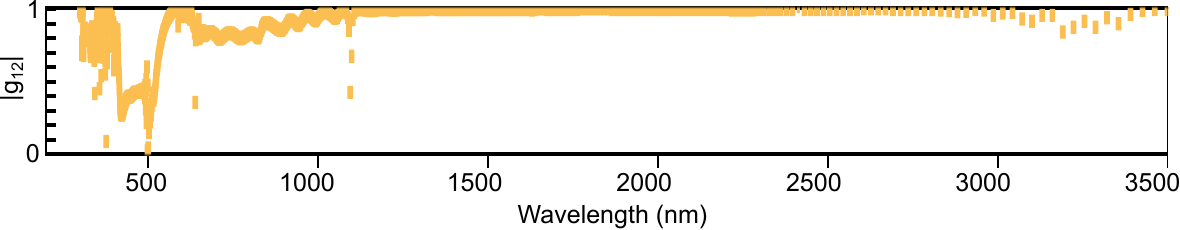}
    \caption{
    Calculated first-order coherence.}
    \label{fig:coherence}
\end{figure}
\FloatBarrier
\bibliography{1Gmain}